\documentclass[aip,jcp,reprint,notitlepage,amsmath,amsfonts,amssymb,floatfix,nolongbibliography,superscriptaddress]{revtex4-2}
\usepackage{graphicx}
\usepackage{bm}

\begin{document}
\title{Strong-Weak Duality via Jordan-Wigner Transformation: Using Fermionic Methods for Strongly Correlated $su(2)$ Spin Systems}
\author{Thomas M. Henderson}
\affiliation{Department of Chemistry, Rice University, Houston, TX 77005-1892}
\affiliation{Department of Physics and Astronomy, Rice University, Houston, TX 77005-1892}

\author{Guo P. Chen}
\affiliation{Department of Chemistry, Rice University, Houston, TX 77005-1892}

\author{Gustavo E. Scuseria}
\affiliation{Department of Chemistry, Rice University, Houston, TX 77005-1892}
\affiliation{Department of Physics and Astronomy, Rice University, Houston, TX 77005-1892}
\date{\today}

\begin{abstract}
The Jordan-Wigner transformation establishes a duality between $su(2)$ and fermionic algebras. We present qualitative arguments and numerical evidence that when mapping spins to fermions, the transformation makes strong correlation weaker, as demonstrated by the Hartree-Fock approximation to the transformed Hamiltonian. This result can be rationalized in terms of rank reduction of spin shift terms when transformed to fermions. Conversely, the mapping of fermions to qubits makes strong correlation stronger, complicating its solution when one uses qubit-based correlators. The presence of string operators poses challenges to the implementation of quantum chemistry methods on classical computers, but these can be dealt with using established techniques of low computational cost. Our proof of principle results for XXZ and J$_1$-J$_2$ Heisenberg (in 1D and 2D) indicate that the JW transformed fermionic Hamiltonian has reduced complexity in key regions of their phase diagrams, and provides a better starting point for addressing challenging spin problems. 
\end{abstract}

\maketitle

\section{Motivation}
With the advent of quantum simulations, a great deal of attention has been paid to using the potential of quantum computers to solve challenging problems in electronic structure theory.\cite{Cau2019,Bauer2020}  One faces an immediate difficulty when attempting to do so, however: electronic structure problems are expressed in terms of fermionic operators which obey anticommutation relations, and quantum computers use qubits which instead have $su(2)$ commutation relations.  One must therefore find a faithful mapping from the one language to the other.  Fortunately, this problem is easily solved, and one can correctly map fermionic Hamiltonians to spin Hamiltonians (${F \to su(2)}$) using the Jordan-Wigner (JW) transformation\cite{Jordan1928} or its variants.\cite{Bravyi2002,Seeley2012}

On the other hand, the inverse transformation (${su(2) \to F}$) applied to the XXZ chain at $\Delta = 0$ is a textbook example\cite{Nishemori2011} of a duality where strongly correlated spins become non-interacting fermions. Inspired by this result, we map this and other spin Hamiltonians over their entire phase diagrams to fermions, and solve them numerically with standard quantum chemistry techniques on a classical computer. Our main result is that across strong interaction regimes, spin correlations become weaker in the fermion frame. In pursuing this work, we are also inspired by Batista and Ortiz\cite{Batista2001,Batista2004}, who have studied general spin-particle connections, including JW, and demonstrated how translating the language can help find simple analytic solutions to problems with intricate interactions.\cite{Batista2001}  Similarly, JW transformation has been applied to, for example, interconvert the one-dimensional Ising model and the Kitaev chain model for p-wave superconductivity.\cite{Kitaev2009,Bardyn2012,Zvyagin2013,Greiter2014}

To the best of our knowledge, a study like the one carried out in this paper has not been presented in the literature (though see, for example, Ref. \onlinecite{Gebhard2022}).  Perhaps the string operators appearing in the JW transformation have acted as a deterrent in classical computations, but despite their many-body character, they can be efficiently manipulated using established tools.\cite{Wahlen-Strothman2015,Wahlen-Strothman2016} Even though exact diagonalization is identical in both frames, the major gain in this approach stems from simple fermion approximations like Hartree-Fock (HF), which correspond to many-body ansatze in the spin frame. Using this idea, we report promising results for reducing the complexity of the strongly correlated spin problems under consideration.

\section{The Jordan-Wigner Transformation}
First proposed by Jordan and Wigner in 1928,\cite{Jordan1928} the JW transformation maps spinless fermionic creation and annihilation operators (i.e. spinorbital operators) to spin raising and lowering operators.  To obtain the correct anticommutation relations, the JW transformation uses extra operators (``Jordan-Wigner strings'') whose sole role is to enforce the proper algebra.  We can thus write
\begin{subequations}
\begin{align}
c_p^\dagger &\to S_p^+ \, \tilde{\phi}_p,
\\
c_p &\to S_p^- \, \tilde{\phi}_p,
\end{align}
\end{subequations}
where the JW strings are
\begin{equation}
\tilde{\phi}_p = \tilde{\phi}_p^\dagger = \prod_{q<p} \mathrm{e}^{\pm \mathrm{i} \, \pi \, \left(S_q^z+1/2\right)} = \prod_{q<p} \left(2 \, S_q^z\right).
\end{equation}
Note that the JW string $\tilde{\phi}_p$ is many body, and commutes with all $S^z$ operators and with the raising and lowering operators $S_q^\pm$ for $q \ge p$.  The important point is that while $S_p^+$ and $S_q^-$ have $su(2)$ commutation relations, $S_p^+ \, \tilde{\phi}_p$ and $S_q^- \, \tilde{\phi}_q$ have fermionic anticommutation relations.

We can invert this transformation to map spin operators to spinless fermions instead:
\begin{subequations}
\begin{align}
S_p^+ &\to c_p^\dagger \, \phi_p,
\\
S_p^- &\to c_p \, \phi_p,
\\
\phi_p &= \prod_{q<p} \mathrm{e}^{\pm \mathrm{i} \, \pi \, n_q} = \prod_{q<p} \, \left(1 - 2 \, n_q\right),
\\
n_p &= c_p^\dagger \, c_p.
\end{align}
\end{subequations}
Using the fact that $[S_p^+,S_p^-] = 2 \, S_p^z$ together with the fermionic anticommutation relationship between $c_p^\dagger$ and $c_p$, and the fact that $\phi_p^2 = 1$, one finds that
\begin{equation}
S_p^z \to n_p - \frac{1}{2} = \bar{n}_p.
\label{Eqn:DefNBar}
\end{equation}

Setting aside the JW strings for a moment, it is important to note that the JW transformation maps two-body spin operators $S_p^+ \, S_q^-$ to one-body fermion operators $c_p^\dagger \, c_q$ and vice-versa.  This suggests that when mapping a fermionic Hamiltonian to qubits, the problem is likely to become more strongly correlated (for which reason we also generally transform the fermionic correlation operators rather than directly using qubit correlation operators, but see, for example, Refs. \onlinecite{Ryabinkin2018,Ryabinkin2018b,Ryabinkin2020}).  On the other hand, when mapping spin systems to fermions, the problem should become less strongly correlated. It is this basic idea we intend to exploit.

Let us therefore map a generic spin Hamiltonian with one- and two-body terms.  We do not assume the spin Hamiltonian has global $S^2$ as a symmetry, but we will assume that it has global $S^z$ symmetry.  This ensures that the transformed fermionic Hamiltonian conserves fermionic particle number.  Of course we can solve fermionic Hamiltonians which do not have number symmetry, but it is not the subject of this work.  A generic spin Hamiltonian of the kind we have discussed can be given by
\begin{align}
H_\mathrm{S} &= H_0 + \sum_p x_p \, S_p^z + \sum_{p < q} \, x_{pq} \, S_p^z \, S_q^z
\label{Def:HSpin}
\\
 &+ \sum_{p < q} J_{pq} \, \left(S_p^+ \, S_q^- + S_q^+ \, S_p^-\right)
 \nonumber
\\
 &+ \mathrm{i} \, \sum_{p<q} K_{pq} \, \left(S_p^+ \, S_q^- - S_q^+ \, S_p^-\right).
\nonumber
\end{align}
By adjusting the coefficients $H_0$, $x_p$, $x_{pq}$, and $J_{pq}$, we can distinguish between various classes of Hamiltonian.  For example, the nearest-neighbor Heisenberg model would set $H_0 = x_p = 0$ and $x_{pq} = 2 \, J_{pq} = J$ for nearest-neighbors $p$ and $q$.  Of course not every $su(2)$ Hamiltonian is of the form given in Eqn. \ref{Def:HSpin}, but many of the most interesting are.  Even for those which are not, one can sometimes first use an $su(2)$ mean-field calculation\cite{Ryabinkin2018b} to transform the $su(2)$ Hamiltonian to adopt the desired form.

We transform this Hamiltonian to fermions as
\begin{align}
H_\mathrm{F} &= H_0 + \sum_p x_p \, \bar{n}_p + \sum_{p < q} x_{pq} \, \bar{n}_p \, \bar{n}_q
\\
 &+ \sum_{p < q} J_{pq} \, \left(c_p^\dagger \, \phi_p \, \phi_q \, c_q + c_q^\dagger \, \phi_q \, \phi_p \, c_p\right)
\nonumber
\\
 &+ \mathrm{i} \, \sum_{p<q} K_{pq} \, \left(c_p^\dagger \, \phi_p \, \phi_q \, c_q - c_q^\dagger \, \phi_q \, \phi_p \, c_p\right).
\nonumber
\end{align}
It will prove convenient in what follows to move the JW strings to one side or the other of the fermionic operators.  As we prove in the appendix, $\phi_q$ and $c_p$ commute (anticommute) when $p \ge q$ ($p<q$), which means that
\begin{equation}
c_p^\dagger \, \phi_p \, \phi_q \, c_q = 
\begin{cases}
 c_p^\dagger \, c_q \, \phi_p \, \phi_q & \qquad p>q,  \\
-c_p^\dagger \, c_q \, \phi_p \, \phi_q & \qquad p<q.
\end{cases}
\end{equation}
Using all of these facts brings us to 
\begin{align}
H_\mathrm{F} &= H_0 + \sum_p x_p \, \bar{n}_p + \sum_{p < q} x_{pq} \, \bar{n}_p \, \bar{n}_q
\\
 &+ \sum_{p < q} J_{pq} \, \left(-c_p^\dagger \, c_q + c_q^\dagger \, c_p\right) \, \phi_p \, \phi_q
\nonumber
\\
 &+ \mathrm{i} \, \sum_{p<q} K_{pq} \, \left(-c_p^\dagger \, c_q - c_q^\dagger \, c_p\right) \, \phi_p \, \phi_q.
\nonumber
\end{align}
Note that the Hamiltonian is in general very high in operator rank, due to the presence of the Jordan-Wigner strings $\phi_p$ and $\phi_q$.  This complicates solving the Hamiltonian, because we would like to start with Hartree-Fock theory and then incorporate correlations via such techniques as coupled cluster theory or configuration interaction.  Nothing about these approaches is conceptually more difficult with a Hamiltonian with more than two-body interactions, but of course the practical realization of these methods relies on the Hamiltonian having low-operator rank.  Even the three-body effective Hamiltonian of nuclear physics renders conventional fermionic methods rather challenging (but possible); a Hamiltonian with, say, a 6-body interaction would be in general intractable.  Fortunately, we have a way forward.

In certain situations, the JW strings disappear, and the difficulty we have just discussed is eliminated.  Indeed, one major claim to fame of the JW mapping is that for the 1D XXZ model, the JW strings do disappear entirely, so that at $\Delta = 0$ the Hamiltonian maps to a simple Hamiltonian for non-interacting fermions.\cite{Nishemori2011} Of course this Hamiltonian is trivially solvable as a fermionic Hamiltonian, even though as a spin model, it is far more complicated to solve (but can be solved via a Bethe ansatz). This is an example of strong-weak duality, in which problems which are strongly correlated in one language become weakly correlated in another.\cite{Batista2004}

In general, however, the JW strings must be accounted for.  In the lattice basis -- that is, the basis of fermionic determinants created by the mapped operators $c_p^\dagger$ -- the action of the JW strings is simple, because lattice-basis determinants are eigenstates of the JW strings.  In practice, the matrix elements of the fermionic Hamiltonian between lattice determinants are the same as those of the spin Hamiltonian between product states in which lattices sites occupied/empty in the fermionic picture correspond to lattice sites with $\uparrow$/$\downarrow$ spins in the spin picture.

We can use this simplicity to do standard Hartree-Fock calculations.  Imagine transforming the Hamiltonian to a new basis via a one-body unitary rotation:
\begin{subequations}
\begin{align}
H_F &\to \mathrm{e}^{-U_1} \, H_F \, \mathrm{e}^{U_1},
\\
U_1 &= \sum t_i^a \, \left(c_a^\dagger \, c_i - c_i^\dagger \, c_a\right)
\end{align}
\end{subequations}
where lattice sites $i$ and $a$ are occupied and empty in the reference determinant.  Hartree-Fock requires us to minimize the energy with respect to the parameters $t_i^a$.  The fermionic operators $c$ and $c^\dagger$ transform to new fermionic operators $a$ and $a^\dagger$:
\begin{equation}
\mathrm{e}^{-U_1} \, c_p^\dagger \, \mathrm{e}^{U_1} = a_p^\dagger = \sum C_{qp} \, c_q^\dagger
\end{equation}
where the coefficient matrix $\mathbf{C}$ is just the exponential of the matrix representation of $U_1$.  The Jordan-Wigner string $\phi_p$ is the exponential of a one-body operator, so upon transformation with $U_1$ it transforms into the exponential of some other one-body operator, i.e. a Thouless rotation.\cite{Thouless1960}  That is,
\begin{subequations}
\begin{align}
\mathrm{e}^{-U_1} \, \phi_p \, \mathrm{e}^{U_1}
 &= \mathrm{e}^{-U_1} \, \mathrm{e}^{\mathrm{i} \, \pi \, \sum_{r<p} c_r^\dagger \, c_r} \, \mathrm{e}^{U_1}
\\
 &= \mathrm{e}^{\mathrm{i} \, \pi \, \sum_{r<p} \mathrm{e}^{-U_1} \, c_r^\dagger \, c_r \, \mathrm{e}^{U_1}}
\\
 &= \mathrm{e}^{\mathrm{i} \, \pi \, \sum_{r<p} a_r^\dagger \, a_r}
\\
 &= \mathrm{e}^{\mathrm{i} \, \pi \, \sum_{r<p} \sum_{st} C_{sr} \, C_{tr}^\star \, c_s^\dagger \, c_t}
\\
 &= \mathrm{e}^{\sum_{st} \mathcal{C}^p_{st} \, c_s^\dagger \, c_t}.
\end{align}
\end{subequations}
Clearly,
\begin{equation}
\mathcal{C}^p_{st} = \mathrm{i} \, \pi \, \sum_{r<p} C_{sr} \, C_{tr}^\star
\end{equation}
is antihermitian:
\begin{equation}
\left(\mathcal{C}^p_{st}\right)^\star = -\mathrm{i} \, \pi \, \sum_{r<p} C_{sr}^\star \, C_{tr} = -\mathcal{C}^p_{ts}.
\end{equation}
Thus, $\phi_p$ transforms into a unitary Thouless transformation, as does $\phi_q$, and their product.  We will generically write
\begin{equation}
\mathrm{e}^{-U_1} \, \phi_p \, \phi_q \, \mathrm{e}^{U_1} = \mathrm{e}^{\mathcal{U}_{pq}}
\end{equation}
where $\mathcal{U}_{pq}$ is an antihermitian one-body operator.

Bearing all this in mind, let us examine the transformation of a single term in the Hamiltonian.  We have
\begin{subequations}
\begin{align}
\mathrm{e}^{-U_1} \, J_{pq} \, c_p^\dagger \, &c_q \, \phi_p \, \phi_q \, \mathrm{e}^{U_1} = J_{pq} \, a_p^\dagger \, a_q \, \mathrm{e}^{\mathcal{U}_{pq}}
\\
 &= \sum_{rs} J_{pq} \, C_{rp} \, C_{sq}^\star \, c_r^\dagger \, c_s \, \mathrm{e}^{\mathcal{U}_{pq}}.
\end{align}
\end{subequations}
The expectation value of this single term is then
\begin{align}
\langle 0| \mathrm{e}^{-U_1} \, J_{pq} \, c_p^\dagger \, &c_q \, \phi_p \, \phi_q \, \mathrm{e}^{U_1} |0\rangle
\nonumber
 \\
  &= \sum_{rs} C_{rp} \, C_{sq}^\star \, J_{pq} \, \langle 0| c_r^\dagger \, c_s \, |\Phi_{pq}\rangle
\end{align}
where
\begin{equation}
|\Phi_{pq}\rangle = \mathrm{e}^{\mathcal{U}_{pq}} |0\rangle
\end{equation}
is a single determinant which is generally not orthogonal to $|0\rangle$.  This matrix element can be evaluated using a generalized version of Wick's theorem.\cite{Balian1969}

Thus, matrix elements of the transformed Hamiltonian can be evaluated as a straightforward if tedious exercise in computing matrix elements between the reference determinant $\langle0|$ and the Thouless-rotated determinant $|\Phi_{pq}\rangle$.  We have, admittedly, a different rotated determinant for every JW string, but there are only $\mathcal{O}(M^2)$ such determinants to construct, where $M$ is the number of lattice sites.  To the extent that we need matrix elements between nonorthogonal Slater determinants, with different transformed determinants for different Hamiltonian terms, all of this bears a practical resemblance to the Lie Algebraic Similarity Transformation (LAST) theory.\cite{Wahlen-Strothman2015,Wahlen-Strothman2016}  Unlike in LAST, though, here we use a unitary transformation, and everything remains variational.

\begin{figure}[t]
\includegraphics[width=\columnwidth]{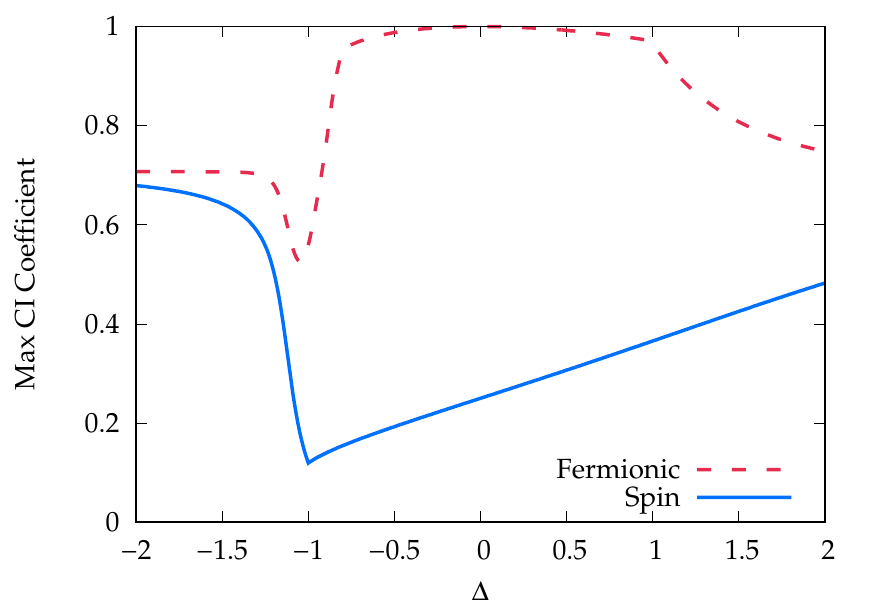}
\caption{Largest CI coefficients in the exact wavefunction for the 8-site XXZ chain with $S^z = 0$.  The fermionic HF changes character near $\Delta = \pm 1$; both fermionic and spin wave functions have CI coefficients which go to $1/\sqrt{2}$ for large $|\Delta|$.
\label{Fig:XXZCoefficients}}
\end{figure}

After transforming the Hamiltonian, configuration interaction (CI) and linearized coupled cluster (LCC) approaches are fairly straightforward, though including the full exponential in coupled cluster (CC) theory is computationally quite challenging and comparable to the obstacles presented by symmetry-projected CC theory.\cite{Qiu2017,Song2022}  The key idea is to transform the correlation operators from the molecular orbital basis to the lattice basis, so that all matrix elements required are simple when the wave function is linear in the correlator.  When the wave function is exponential in the correlator, this problem is generally intractable, and the same sorts of techniques used in symmetry-projected coupled cluster theories would be required.

In order to present proof of principle results, for now we circumvent these problems by working with a fermionic full configuration interaction (FCI) code.  We build the Slater determinant matrix representation of the fermionic Hamiltonian in the lattice basis.  We do a Hartree-Fock calculation by repeated CI singles (CIS).  That is, we write the wave function as ${|\Psi\rangle = \left(1 + \sum t_i^a \, c_a^\dagger \, c_i\right) |0\rangle}$ and minimize the energy with respect to $t_i^a$.  Then we build the FCI matrix representation of the operator $T_1 = \sum t_i^a \, c_a^\dagger \, c_i$ and exponentiate $T_1 - T_1^\dagger$, which we use to transform the FCI matrix to a new basis.  We then repeat this procedure until we converge to $t_i^a = 0$ at which point the reference determinant $|0\rangle$ is a Hartree-Fock determinant, with well-defined occupied and virtual single-particle levels.  With these in hand, we can carry out truncated configuration interaction calculations using the FCI matrix in the molecular orbital basis.

\begin{figure}[t]
\includegraphics[width=\columnwidth]{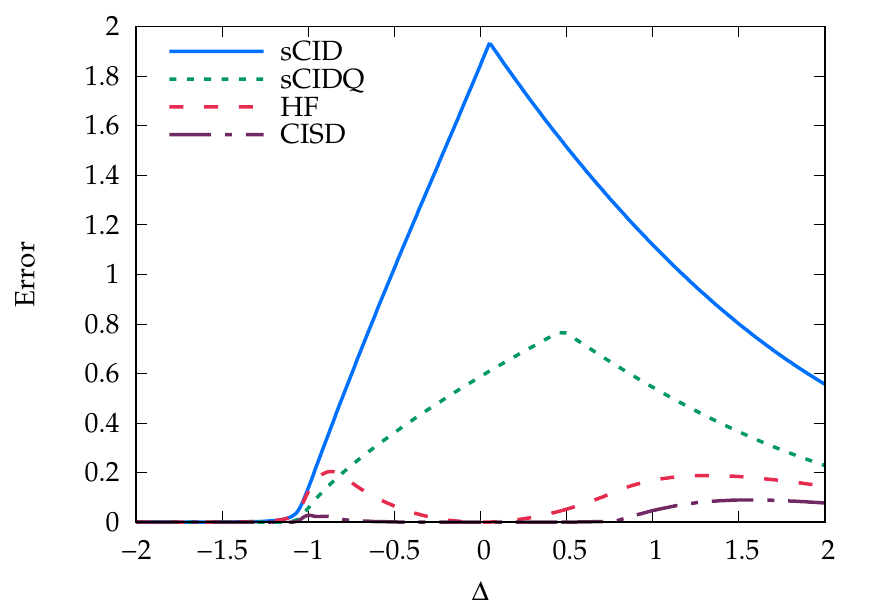}
\caption{Energy error for the 12-site XXZ chain with $S^z = 0$.  All methods become energetically exact as $|\Delta|$ becomes large.
\label{Fig:XXZEnergyErrors}}
\end{figure}

\section{Results}
\subsection{The XXZ Hamiltonian}
The nearest-neighbor XXZ Hamiltonian writes
\begin{equation}
H_S = \sum_{\langle pq \rangle} \, \left(\frac{1}{2} \, S_p^+ \, S_q^- + \frac{1}{2} \, S_q^+ \, S_p^- + \Delta \, S_p^z \, S_q^z\right)
\end{equation}
where the notation $\langle pq \rangle$ means we include only sites $pq$ adjacent in the lattice.  We will consider both a one-dimensional (1D) lattice and a two-dimensional (2D) rectangular lattice.

In 1D, as we have noted earlier, the JW strings drop out entirely if we have open boundary conditions (OBC).  With periodic boundary conditions, there is a single JW string when sites 1 and $M$ couple, but one which can be handled without too much difficulty.  Without JW strings to worry about, the $S^+ \, S^-$ part of the Hamiltonian becomes a one-body operator $c^\dagger \, c$, and while the $S^z \, S^z$ term becomes two-body, transforming as it does to $\bar{n} \, \bar{n}$, it vanishes at $\Delta = 0$.  Accordingly, the fermionic HF is exact at $\Delta = 0$, even though from the perspective of spin configurations, this area is strongly correlated.  We can see that spins are strongly correlated near $\Delta = 0$ simply by noting that at $\Delta = 0$, all spin configurations are energetically degenerate.

Figure \ref{Fig:XXZCoefficients} shows the largest CI coefficient in the exact wave function for the half-filled ($S^z = 0$) 8-site, 1D XXZ Hamiltonian with OBC as a function of $\Delta$.  Weak correlation implies that this coefficient is close to 1.  As $\Delta \to \infty$ the system adopts a N\'eel configuration; this configuration is doubly-degenerate so the exact CI coefficients approach $1/\sqrt{2}$ in magnitude.  As $\Delta \to -\infty$ the system places the $\uparrow$ spins all on one side or the other.  This configuration is again doubly-degenerate and the exact CI coefficients again approach $1/\sqrt{2}$.  The fermionic Hamiltonian retains these features for large $|\Delta|$, but where for modest $\Delta$ the spin Hamiltonian has no dominant configuration, the fermionic Hamiltonian is dominated by a single determinant for $|\Delta| \lesssim 1$.  All of this is exactly as we would expect.  The same basic features persist for larger chains, but the ground state for $\Delta \lesssim -1$ becomes very nearly degenerate and without symmetry-adapting the Hamiltonian, even exact diagonalization tends to produce ground state wave functions which break symmetry when one works, as we do, in double precision.  We note in passing that any strong correlations deriving from the CI coefficients approaching $1/\sqrt{2}$ for large $|\Delta|$ can, in fermionic language, readily be handled by lattice symmetry projection at minimal additional cost.\cite{PHF}

\begin{figure}[t]
\includegraphics[width=\columnwidth]{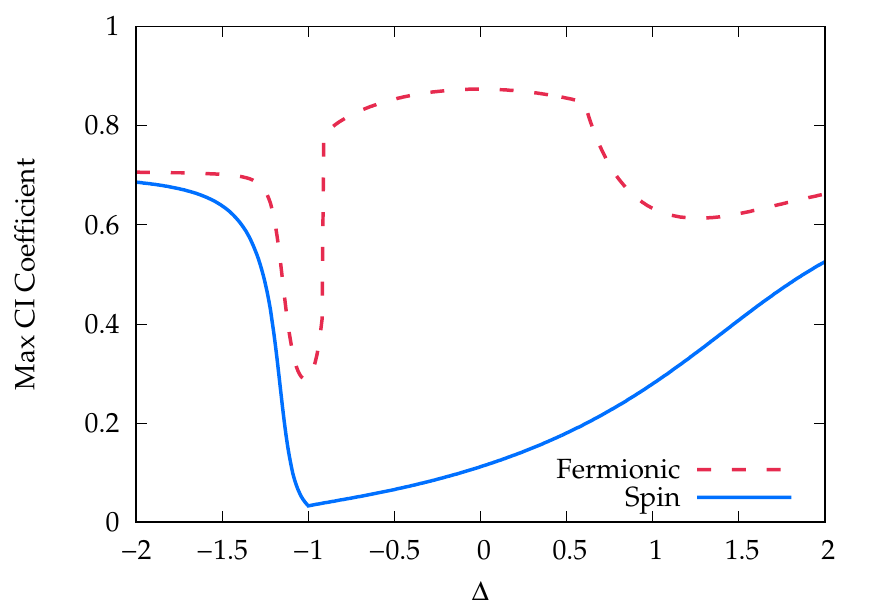}
\\
\includegraphics[width=\columnwidth]{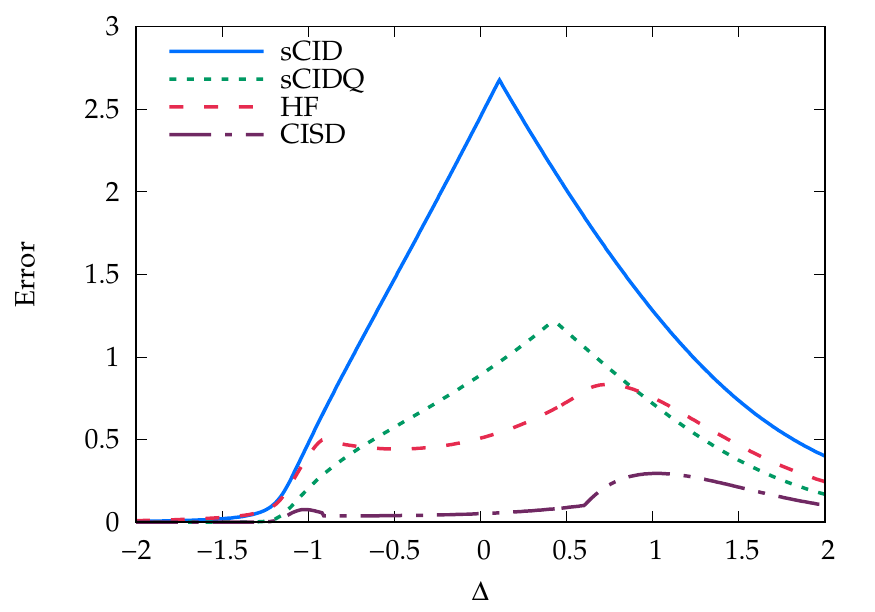}
\caption{Top: Largest CI coefficients in magnitude for the $2 \times 6$ XXZ system with a square lattice, both in spin and fermionic language.  Bottom: Energy errors in the same system.  Again, all methods become energetically exact as $|\Delta|$ becomes large.
\label{Fig:XXZ2D}}
\end{figure}

Note that while looking at degeneracy of the spin configurations suggests that the strongest correlations occur near $\Delta = 0$ in the spin picture, the spin wave function is in fact most multiconfigurational at $\Delta = -1$, where the antisymmetrized geminal power is exact and all configurations have equal coefficients in absolute value.  The fermionic wave function is also most strongly correlated in the vicinity of $\Delta = -1$.

Now we add correlation by configuration interaction methods; for now we forego CC methods because, while we can implement them in our FCI code, it is not completely clear how one would implement fermionic CC methods in practice for reasons we have already discussed.  The fermionic CI with singles and doubles (CISD) is presumably familiar to the reader.  In the spin case we use a spin CI with double excitations (sCID), for which
\begin{equation}
|\Psi\rangle = \left(1 + \sum_{ia} c_i^a \, S_a^+ \, S_i^-\right) |0\rangle_S
\end{equation}
where sites $i$ have $\uparrow$ spins in the reference and sites $a$ have $\downarrow$ spins in the reference $|0\rangle_S$, which we take to be one of the lowest-energy spin configurations.  In fermionic language, this would correspond to a CIS wave function dressed by JW phases, acting in the lattice basis rather than in the HF orbital basis.  To make a fairer comparison to the fermionic CISD, we also implement a spin CI with both double and quadruple excitations (sCIDQ), which has the same $\mathcal{O}(M^4)$ number of excitation operators as does fermionic CISD.  Results appear in Fig. \ref{Fig:XXZEnergyErrors}.  Clearly the fermionic HF calculation is almost everywhere superior even to the spin CIDQ.  Of course CISD only improves the situation further.  The improvement is particularly notable for $|\Delta| \lesssim 1$, which makes sense since the HF is exact at $\Delta = 0$.

\begin{figure}[t]
\includegraphics[width=0.5\columnwidth]{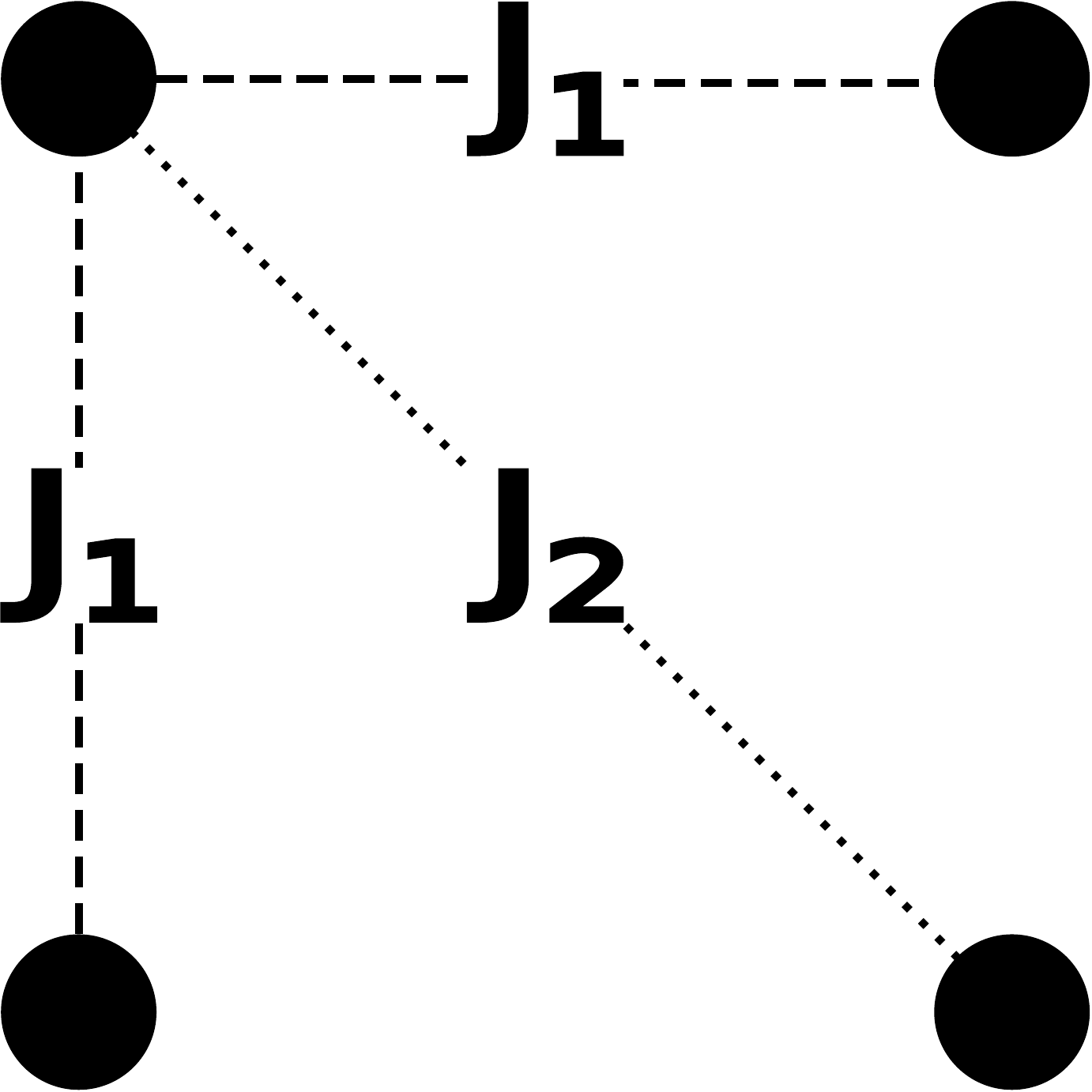}
\caption{Schematic illustration of a 2D square $J_1-J_2$ lattice, where $J_1$ couples nearest neighbors and $J_2$ couples next-nearest neighbors.
\label{Fig:J1J2}}
\end{figure}

One may be concerned that we are relying on the exactness of HF at $\Delta = 0$ to achieve accurate results.  To see that this is not the case, Fig. \ref{Fig:XXZ2D} shows results for the 2$\times$6 2D XXZ system with a rectangular lattice, again with open boundary conditions and $S^z = 0$.  This lattice is quasi-one--dimensional, and JW strings now appear in the fermionic Hamiltonian.  As such, Hartree-Fock is not exact at $\Delta = 0$, but we still see that there is a single determinant which dominates there, where in the spin frame the system has no clear important spin configuration.  This is reflected in the energetic errors.  Of course a true 2D XXZ lattice would require more sites in each direction, but that is unfortunately beyond the reach of an implementation based off of an FCI code and will be addressed in future work.

\begin{figure}[t]
\includegraphics[width=\columnwidth]{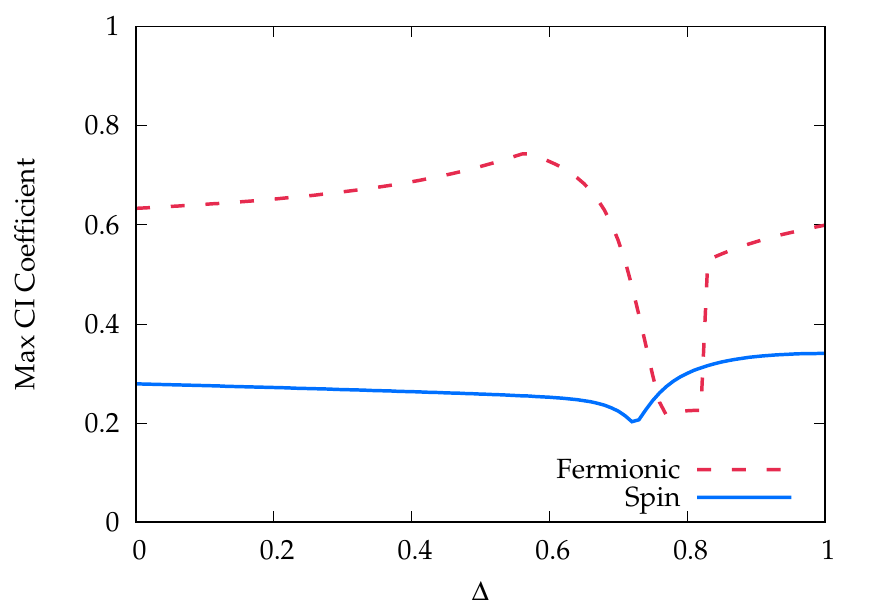}
\\
\includegraphics[width=\columnwidth]{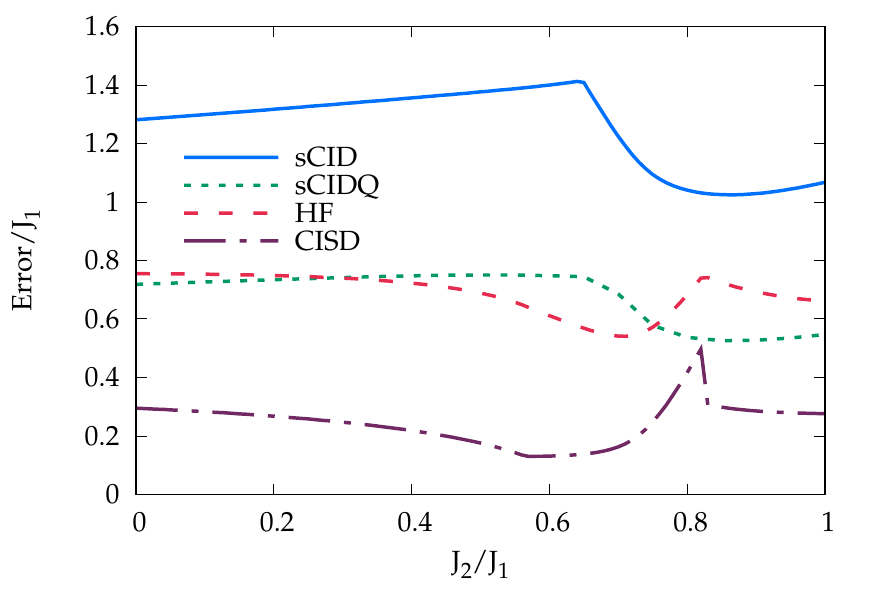}
\caption{Top: Largest CI coefficients in magnitude for the $2 \times 6$ $J_1-J_2$ system with a square lattice, both in spin and fermionic language.  Bottom: Energy errors in the same system.  
\label{Fig:J1J2Results}}
\end{figure}

\subsection{The $J_1-J_2$ Hamiltonian}
We briefly consider a different 2D Hamiltonian, the $J_1-J_2$ Heisenberg model, given by
\begin{equation}
H_S = J_1 \, \sum_{\langle pq \rangle} \vec{S}_p \cdot \vec{S}_q + J_2 \, \sum_{\langle\langle pq \rangle \rangle} \vec{S}_p \cdot \vec{S}_q.
\end{equation}
Here, $\langle \langle pq \rangle \rangle$ denotes sites $p$ and $q$ which are next-nearest neighbors, i.e. sites which are displaced by 1 from each other in both the $x$ and $y$ directions of the lattice (see Fig. \ref{Fig:J1J2}).  For small $J_2$, this Hamiltonian adopts a Ne\'el configuration in which adjacent sites have antiparallel spins, while for large $J_2$ a striped phase is preferred where all sites along a single row (or a single column) have parallel spins, and adjacent rows (or columns) are antiparallel.  For intermediate $J_2$, the nature of the state is much less certain.

\begin{figure*}[t]
\includegraphics[width=0.3 \textwidth]{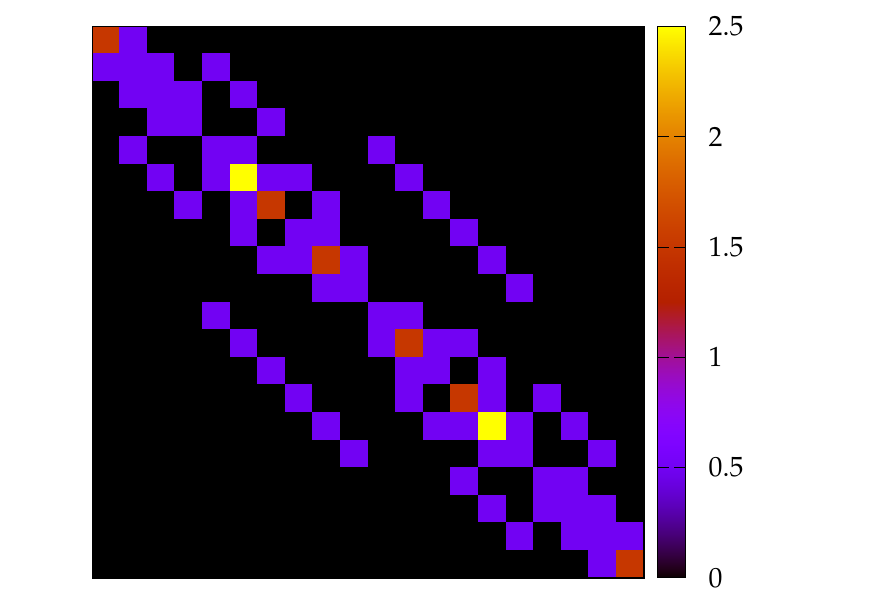}
\hfill
\includegraphics[width=0.3 \textwidth]{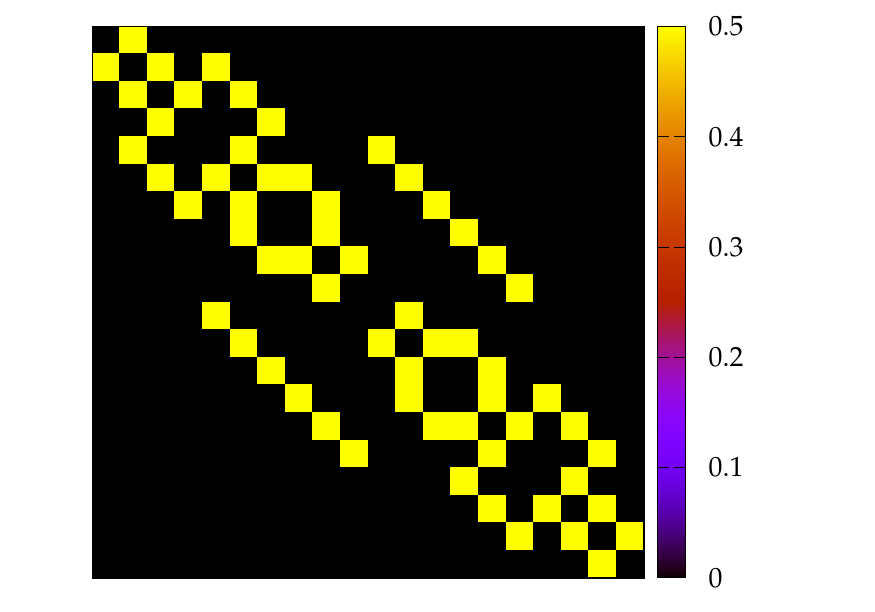}
\hfill
\includegraphics[width=0.3 \textwidth]{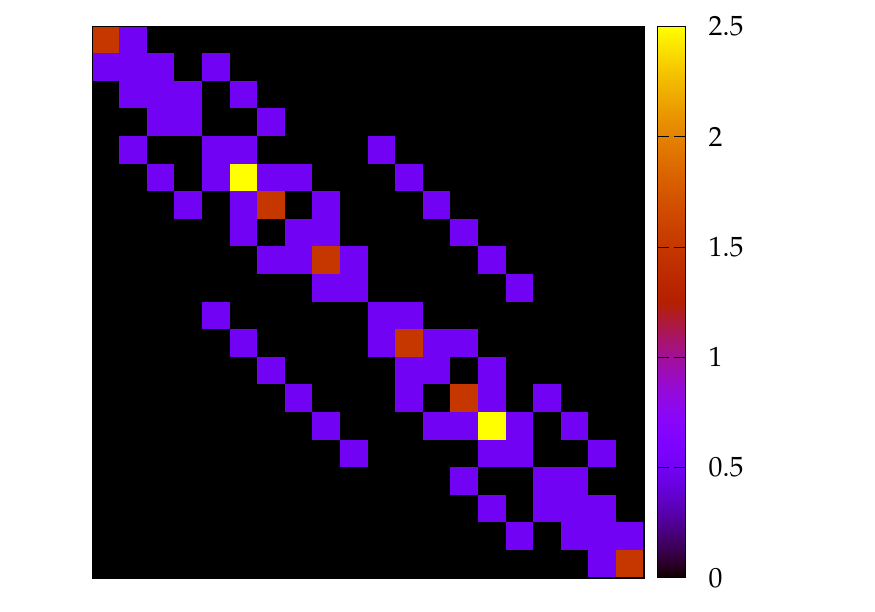}
\\
\includegraphics[width=0.3 \textwidth]{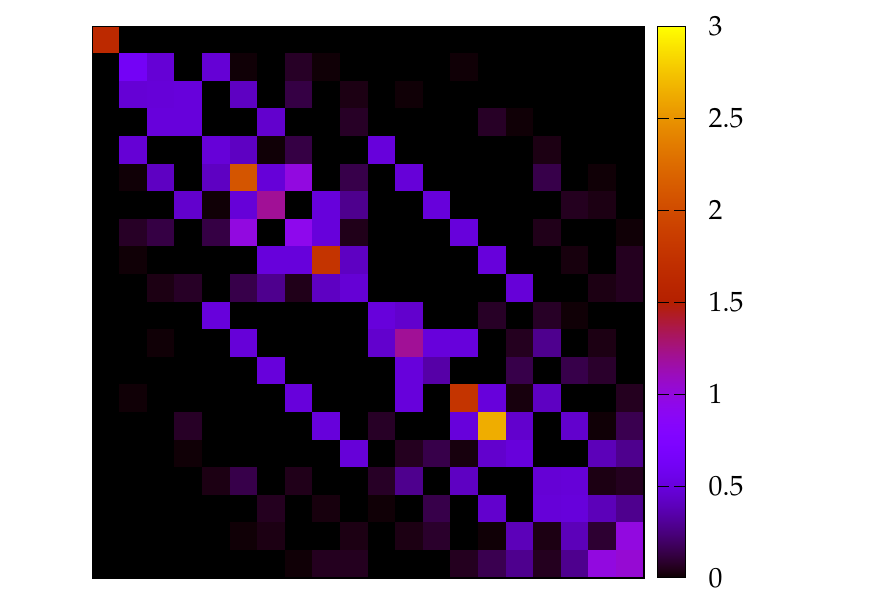}
\hfill
\includegraphics[width=0.3 \textwidth]{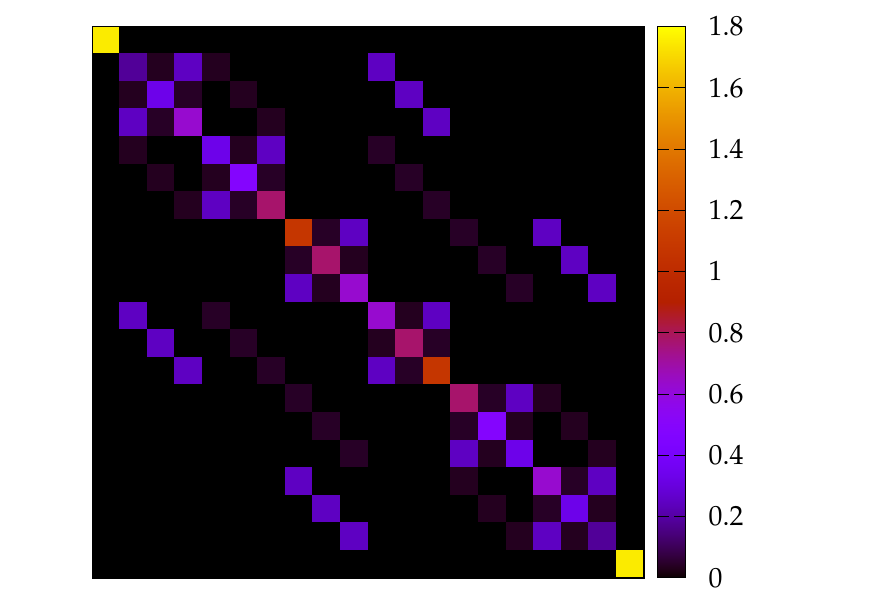}
\hfill
\includegraphics[width=0.3 \textwidth]{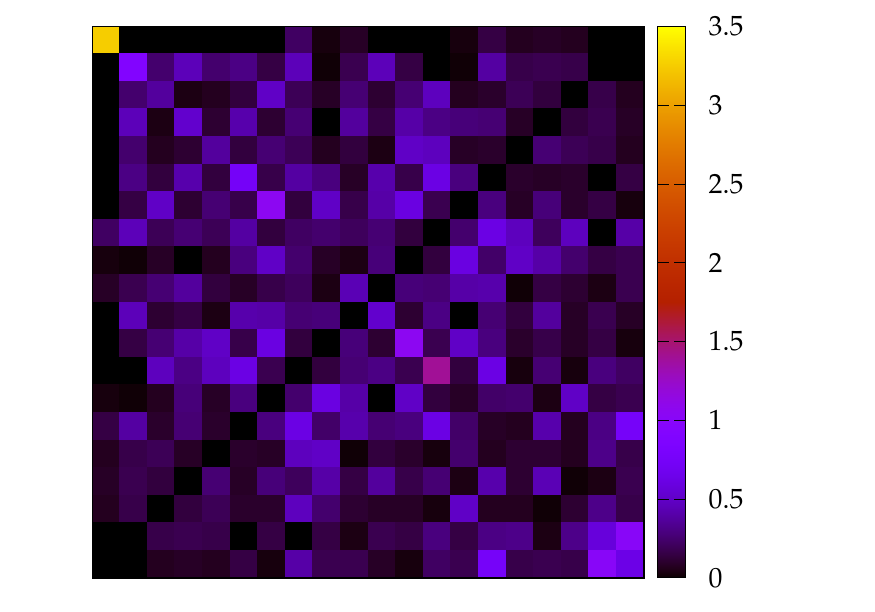}
\caption{Hamiltonian matrices, in absolute value, for the half-filled 6-site XXZ chain.  Top row: Matrices in the lattice basis.  Bottom row: Matrices in the Hartree-Fock basis.  From left to right the columns are at $\Delta = -2$, $\Delta = 0$, and $\Delta = 2$.  Note that the lattice-basis Hamiltonian depends on $\Delta$ only in the diagonal, and in absolute value is identical at $\Delta = \pm 2$.
\label{Fig:Matrices}}
\end{figure*}

We show results for the $2 \times 6$ $J_1-J_2$ lattice with open boundary conditions in Fig. \ref{Fig:J1J2Results}.  Again, even Hartree-Fock is roughly equivalent to spin with four-body correlations, and adding correlation to the fermionic treatment is better yet.  There is a discontinuity in the fermionic treatment near $J_2/J_1 \sim 0.8$ which we have been unable thus far to resolve, arising from a transition in the nature of the fermionic state.  One can see the effects of this transition in the CI coefficients of the wave function.  We have shown open boundary conditions simply because the fermionic treatment is exact at $J_2 / J_1 = 1/2$ for $2 \times n$ spin ladders with periodic boundary conditions, despite the nontrivial appearance of JW strings.  This is presumably because $J_2/J_1 = 1/2$ corresponds to the Majumdar-Ghosh point,\cite{Majumdar1969} where the exact wave function has a simple structure.  Because we are able to obtain the exact result at this point, results for the periodic system artificially favor the fermionic treatment.  We do not expect the fermionic treatment to remain exact at $J_2/J_1 = 1/2$ for larger periodic lattices.

\section{Discussion}
Our results show that already, Hartree-Fock in the fermionic picture can give reasonably accurate results for the spin models we have discussed here, and adding correlation with traditional quantum chemistry techniques can lead to exceptional accuracy.  It is worth emphasizing that even the fermionic Hartree-Fock is, in terms of spin operators, a very complicated wave function.  Recall that in practice the fermionic Hartree-Fock wave function is
\begin{equation}
|\mathrm{HF}\rangle = \mathrm{e}^{\sum t_i^a \, \left(c_a^\dagger \, c_i - c_i^\dagger \, c_a\right)} |0\rangle_F
\end{equation}
where $|0\rangle_F$ is a reference determinant in the lattice basis with sites $i$ occupied and $a$ empty.  Translating this to \textit{\textit{su(2)}} language gives us
\begin{equation}
|\mathrm{HF}\rangle \to \mathrm{e}^{\sum t_i^a \, \left(S_a^+ \, \tilde{\phi}_a \, \tilde{\phi}_i \, S_i^- -S_i^+ \, \tilde{\phi}_i \, \tilde{\phi}_a \, S_a^-\right)} |0\rangle_S
\end{equation}
where $|0\rangle_S$ is a spin product state with $\uparrow$ spins in sites $i$ and $\downarrow$ spins in sites $a$.  Without the JW strings, this is a unitary coupled cluster doubles wave function and is, on a classical computer, already intractable.  Adding the JW strings only makes it more complicated, but in the fermionic frame this wave function is straightforward to construct with the only difficulty being the optimization of the parameters $t_i^a$.

To see why this works, it may be helpful to consider the Hamiltonian matrix directly.  We create the matrix representation of an $su(2)$ Hamiltonian by using, as a basis, simple product states in which we begin from a spin vacuum with all sites having $\downarrow$ spin, then act $N$ distinct $S^+$ operators to create an $S^z$ eigenstate with the appropriate eigenvalue.  These $su(2)$ states map directly to fermionic single determinants created by acting the fermionic creation operators $c^\dagger$ on the fermionic vacuum.  Consequently, the matrix representations of the $su(2)$ Hamiltonian and of the fermionic Hamiltonian are identical.  Fermions, however, have the advantage that linear combination of fermionic operators $a_p^\dagger = \sum U_{pq} \, c_q^\dagger$ are themselves properly fermionic when the matrix of coefficients $U_{pq}$ is unitary (i.e. $\{a_p^\dagger,a_q\} = \delta_{pq}$ and $\{a_p^\dagger,a_q^\dagger\} = 0$).  The same is not true for $su(2)$ operators, for which we can mix $S_p^+$, $S_p^z$, and $S_p^-$ but not $S_p^+$ and $S_q^+$ while maintaining $su(2)$ commutation rules, a point also emphasized in Ref. \onlinecite{Ryabinkin2018b}.  The main reason to work with a fermionic representation is that we can take advantage of this unitary transformation to simplify the calculations.

Thus, Fig. \ref{Fig:Matrices} shows Hamiltonian matrices for a small (6 site) XXZ chain at half filling.  The top row shows results for the spin Hamiltonian, before JW mapping.  The matrices in the fermionic basis after JW mapping are identical, by design.  Unlike with spins, however, fermions readily allow Hartree-Fock transformation, and the bottom row shows the same Hamiltonians after Hartree-Fock transformation.  It is apparent that although the Hartree-Fock transformation does not necessarily simplify the description of the excited states, it generally simplifies the ground state, as emphasized by the relative sparsity of the first row and column of the matrix (recall that if the off-diagonal elements of the first row and column vanish, one has of course found an eigenstate of the Hamiltonian).

\begin{figure*}[t]
\includegraphics[width=\columnwidth]{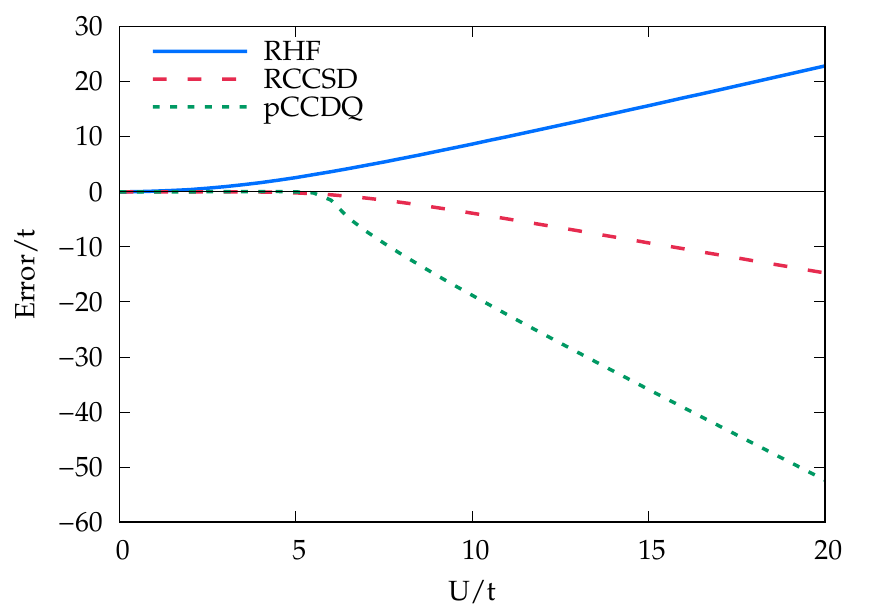}
\hfill
\includegraphics[width=\columnwidth]{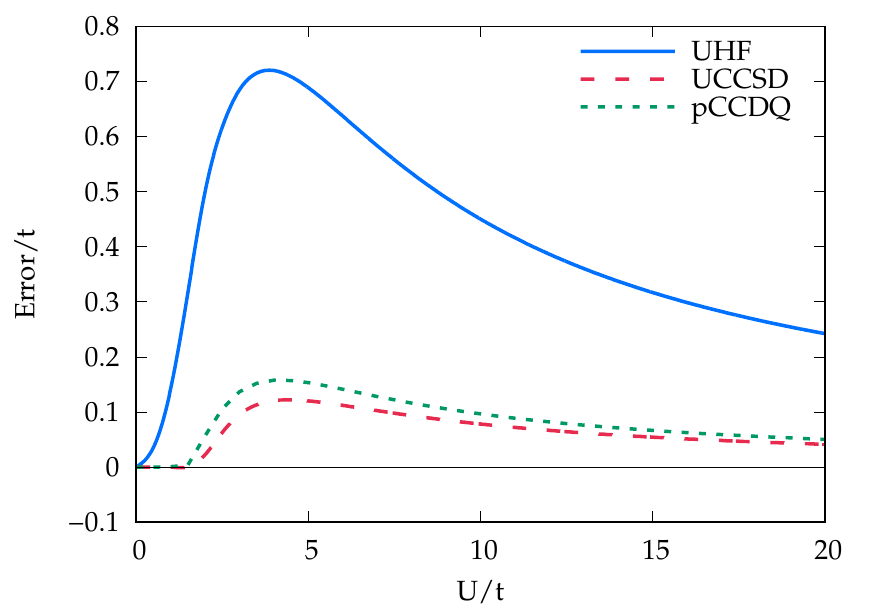}
\caption{Energy errors in the half-filled 6-site Hubbard model.  Left panel: JW mapping done in the RHF basis with periodic boundary conditions.  Right panel: JW mapping done in the UHF basis with open boundary conditions.
\label{Fig:Hubbard}}
\end{figure*}

It is worth emphasizing at this point that the practical benefits of the JW transformation do not obviously extend in the other direction.  That is, we can map fermionic Hamiltonians to qubit Hamiltonians, but while $su(2)$ problems of the sort considered here map to fermionic problems which, after an appropriate mean-field transformation, are more weakly correlated, the reverse is not necessarily true.

To see this, we briefly consider results for the Hubbard model.  In the lattice basis, this Hamiltonian is given by
\begin{equation}
H = -t \, \sum_{\langle pq \rangle} \sum_\sigma \left(c_{p,\sigma}^\dagger \, c_{q,\sigma} + h.c.\right) + U \, \sum_p n_{p,\uparrow} \, n_{p,\downarrow}.
\end{equation}
We first solve this Hamiltonian at the mean-field level, then transform it using JW to a Hamiltonian of $su(2)$ operators; the resulting $su(2)$ Hamiltonian appears in Appendix \ref{App:Mapping}.  Figure \ref{Fig:Hubbard} shows results when the Hamiltonian is solved using restricted Hartree-Fock (RHF) in which the mean-field preserves $S^2$ symmetry, as well as unrestricted Hartree-Fock (UHF) in which it preserves $S^z$ symmetry but not $S^2$.  We treat the resulting transformed Hamiltonian with standard pair coupled cluster\cite{Limacher2013,Limacher2014,Tecmer2014,Boguslawski2014,Stein2014,Henderson2014b} with doubles and quadruples (pCCDQ), in which we write
\begin{subequations}
\begin{align}
|\Psi\rangle &= \mathrm{e}^{T_2 + T_4} |0\rangle,
\\
T_2 &= \sum_{ia} t_i^a \, S_a^+ \, S_i^-,
\\
T_4 &= \frac{1}{4} \, \sum_{ijab} t_{ij}^{ab} \, S_a^+ \, S_b^+ \, S_i^- \, S_j^-
\end{align}
\end{subequations}
and then solve traditional coupled cluster equations:
\begin{subequations}
\begin{align}
E &= \langle 0| \bar{H} |0\rangle,
\\
0 &= \langle 0| S_i^+ \, S_a^- \, \bar{H} |0\rangle,
\\
0 &= \langle 0| S_i^+ \, S_j^+ \, S_a^- \, S_b^- \, \bar{H} |0\rangle,
\\
\bar{H} &= \mathrm{e}^{-\left(T_2 + T_4\right)} \, H \, \mathrm{e}^{T_2 + T_4}.
\end{align}
\end{subequations}
This is roughly equivalent to fermionic coupled cluster with single and double excitations (CCSD).  When the JW mapping is done in the UHF basis, pCCDQ is roughly equivalent to but slightly worse than the corresponding fermionic method.  In the RHF basis, neither approach is particularly good, but the JW-mapped technique breaks down even more severely than does the fermionic approach.  Additionally, the JW-transformed Hamiltonian is not obviously invariant to fermionic occupied-occupied or virtual-virtual rotations, so that we are not entirely convinced that pCCDQ gives unique results.  All of this argues that while practical considerations may require one to treat JW-transformed fermionic Hamiltonians as if they were actually spin Hamiltonians from the beginning, care must be taken because the JW-transformation may convert a less strongly-correlated fermionic Hamiltonian into a more strongly-correlated $su(2)$ Hamiltonian.

\section{Conclusions}
We believe that mapping spin systems to their fermionic counterpart and using fermionic methods to solve them is an underappreciated technique.  Let us reiterate why, conceptually, one may wish to do this.

Spin configurations are eigenstates of the part of the spin Hamiltonian which depends only on local $S^z$ operators.  The local raising and lowering operators $S^+$ and $S^-$ act to create interactions. The more important these interaction terms are compared to the $S^z$ terms, the stronger the correlation when expressed in the language of spins.  However, when transformed to fermions, the two-body interaction $S^+ \, S^-$ maps to a one-body fermionic term $c^\dagger \, c$ together with JW strings.

While the JW strings are many-body, they have the special property that lattice determinants are their eigenstates.  The result is that the interaction part of the Hamiltonian becomes one-body--like in character in the lattice basis (and indeed, for nearest-neighbor 1D systems, the interaction part of the Hamiltonian is strictly one-body).  The fermionic Hamiltonian also has one-body and two-body contributions from the $S^z$ terms, and Hartree-Fock finds the best compromise description, treating all the Hamiltonian terms on a loosely equal footing.  After a Hartree-Fock calculation and corresponding Hamiltonian transformation, one is frequently left with a Hamiltonian which is not too difficult to describe in the language of fermions, even though the original spin Hamiltonian was far from simple to solve. Again, this is a form of strong-weak duality.  

One final point we wish to make is that spin configurations in $su(2)$ map directly to fermionic determinants in the lattice basis, so the spin Hamiltonian matrix and the fermionic Hamiltonian matrix, when the latter is expressed in the lattice basis, are identical.  Accordingly, so are their eigenvectors.  A truncated spin CI is then exactly equivalent to a truncated fermionic CI in the lattice basis (though note that one can do spin coupled cluster or fermionic coupled cluster in the lattice basis and get different results, and the latter is generally superior in our experience).  The key idea here is that we can use Hartree-Fock to transform the fermionic Hamiltonian to a form more amenable to correlation.  Essentially, the lattice basis is a kind of ``atomic orbital basis'' and while a truncated CI can be done in this basis, it is of course generally better to use some suitable molecular orbital basis instead.  Similar ideas can be done directly in terms of spin configurations, but it is simpler when expressed in the language of fermions.

To summarize, the two algebraic representations yield identical results when dealing with exact eigenstates. Our main quantitative finding is that for spin Hamiltonians of the kind given in Eqn. \ref{Def:HSpin}, mean-field theory in the fermion frame, which corresponds to a two-body exponential in the spin representation, is significantly more accurate than mean-field theory in the spin frame. As such, it is a much better starting point for finding accurate solutions, while roughly conserving the computational cost. The improvement is more discernible in the critical region of strongly correlated spin systems, thus truly representing a strong-weak duality.

\section{Data Availability}
 The data that support the findings of this study are available from the corresponding author upon reasonable request.

\begin{acknowledgments}
This work was supported by the U.S. Department of Energy, Office of Basic Energy Sciences.  The Jordan-Wigner aspects were supported under Award DE-SC0019374, and the strong correlation aspects under Award DE-FG02-09ER16053.  G.E.S. is a Welch Foundation Chair (C-0036) and acknowledges useful comments by Gerardo Ortiz and Jorge Dukelsky.
\end{acknowledgments}

\appendix
\section{Commuting Jordan-Wigner Strings and Fermion Operators}
Here, we wish to demonstrate that $\phi_q$ and $c_p$ commute when $p \ge q$, and anticommute when $p<q$.

Recall first that
\begin{equation}
\phi_q = \prod_{r<q} \left(1 - 2 \, n_r\right) = \prod_{r<q} f_r
\end{equation}
where we have defined
\begin{equation}
f_r = 1 - 2 \, n_r.
\end{equation}
Because
\begin{subequations}
\begin{align}
n_p \, c_p &= c_p^\dagger \, c_p \, c_p = 0,
\\
c_p \, n_p &= c_p \, c_p^\dagger \, c_p = \left(1 - c_p^\dagger \, c_p\right) \, c_p = c_p,
\end{align}
\end{subequations}
we see that $f_p$ and $c_p$ anticommute:
\begin{equation}
\{f_p,c_p\}
 = 2 \, c_p - 2 \, c_p \, n_p - 2 \, n_p \, c_p = 0.
\end{equation}
Of course $f_r$ and $c_p$ commute when $r \ne p$.

With all of this in mind, we see that if $p < q$, then
\begin{subequations}
\begin{align}
\phi_q \, c_p
 &= \prod_{r < q} f_r \, c_p
\\
 &= f_p \, c_p \, \prod_{r < q \atop r \ne p} f_r
\\
 &= -c_p \, f_p \, \prod_{r < q \atop r \ne p} f_r
\\
 &= -c_p \, \phi_q.
\end{align}
\end{subequations}
If, on the other hand, $p \ge q$, then $\phi_q$ does not contain $f_p$, so we do not pick up the minus sign.  Put differently, for $p \ge q$, $c_p$ commutes with every term in the product over $r$ which defines $\phi_q$; for $p < q$, $c_p$ commutes with every term but one, with which it instead anticommutes.

Taken together, this means that $\phi_q$ and $c_p$ commute for $p \ge q$ and anticommute for $p<q$.

\section{Fermionic Hamiltonian Mapped to $su(2)$
\label{App:Mapping}}
Suppose we have a fermionic Hamiltonian expressed in the spinorbital basis as
\begin{equation}
H_F = \sum h_{pq} \, c_p^\dagger \, c_q + \frac{1}{4} \, \sum v_{pq,rs} \, c_p^\dagger \, c_q^\dagger \, c_s \, c_r
\end{equation}
where $h_{pq}$ are one-electron integrals and $v_{pq,rs}$ are antisymmetrized two-electron integrals.  Mapped to the spin basis, we obtain
\begin{widetext}
\begin{align}
H_S
  &= \left(
    \frac{1}{4} \, \sum_p h_{pp} \, \bar{S}_p^z
    + \sum_{p < q} h_{pq} \, S^+_p \, S_q^- \, \tilde{\phi}_p \, \tilde{\phi}_q
    + \frac{1}{8} \, \sum_{p < q} v_{pq,pq} \, \bar{S}_p^z \, \bar{S}_q^z
  \right.\\
  \nonumber
  &\qquad + \frac{1}{2} \sum_{p < q < r}
  \left(
    - v_{pq,qr} \, S^+_p \, \bar{S}^z_q \, S^-_r \tilde{\phi}_{p} \, \tilde{\phi}_r
    + v_{pq,pr} \, S^+_q \, \bar{S}^z_p \, S^-_r \tilde{\phi}_q \, \tilde{\phi}_r
    + v_{pr,qr} \, S^+_p \, \bar{S}^z_r \, S^-_q \tilde{\phi}_p \, \tilde{\phi}_q
  \right)\\
  \nonumber
  &\qquad \left.
    + \sum_{p < q < r < s}
    \left(
      - v_{pq,rs} \, S^+_p \, S_q^+ \, S_r^- \, S_s^-
      + v_{pr,qs} \, S^+_p \, S_r^+ \, S_q^- \, S_s^-
      - v_{qr,ps} \, S^+_q \, S^+_r \, S_p^- \, S_s^-
    \right) \tilde{\phi}_p \, \tilde{\phi}_q \, \tilde{\phi}_r \, \tilde{\phi}_s
  \right) + h.c.
\end{align}
\end{widetext}
Here, we have defined
\begin{equation}
\bar{S}_p^z = 2 \, S_p^z + 1,
\end{equation}
in analogy with Eqn. \ref{Eqn:DefNBar}.  This gives 0 or 2 for a site with $\downarrow$ or $\uparrow$ spin.  Note that even setting aside the JW strings, the Hamiltonian is 4-body in terms of spins even though only two-body in terms of fermions.

\bibliography{JordanWignerBib}
\end{document}